\documentclass[journal]{IEEEtran}
\usepackage[protrusion=true,expansion=true]{microtype}

\usepackage[pdftex]{graphicx}
\graphicspath{{../Media/}}
\DeclareGraphicsExtensions{.pdf}

\usepackage[cmex10]{amsmath}
\usepackage{amssymb}
\usepackage{algpseudocode}
\usepackage{algorithm}

\usepackage[caption=false,font=footnotesize]{subfig}

\usepackage{dblfloatfix}

\usepackage{cite}

\usepackage{url}

\newtheorem{proposition}{Proposition}[section]
\newtheorem{assumption}{Assumption}

\newtheorem{theorem}{Theorem}[section]

\newtheorem{problem}{Problem}
\newtheorem{remark}{Remark}[section]

\newcommand{\eqb}[1]{\begin{equation}\label{#1}}
\newcommand{\eqe}{\end{equation}}
\newcommand{\mb}[1]{\mathbf{#1}}
\newcommand{\mbb}[1]{\mathbb{#1}}
\newcommand{\mc}[1]{\mathcal{#1}}

\newcommand{\lp}{\left(}
\newcommand{\rp}{\right)}

\newcommand{\st}{\,|\,}
\newcommand{\fa}{\;\forall \;}

\newcommand{\ib}{\begin{itemize}}
\newcommand{\ie}{\end{itemize}}
\providecommand{\abs}[1]{\lvert#1\rvert}
\providecommand{\norm}[1]{\lVert#1\rVert}

\newsavebox{\ieeealgbox}

\renewcommand{\figurename}{Fig.\;}

\begin{document}

\title{Route Swarm:  Wireless Network\\Optimization through Mobility}

\author{Ryan~K.~Williams,~\IEEEmembership{Student Member,~IEEE}, Andrea~Gasparri,~\IEEEmembership{Member,~IEEE}, and Bhaskar~Krishnamachari,~\IEEEmembership{Member,~IEEE}%
\thanks{R.~K.~Williams and B.~Krishnamachari are with the Department of Electrical Engineering at the University of Southern California, Los Angeles, CA 90089 USA (rkwillia@usc.edu; bkrishna@usc.edu).}%
\thanks{A.~Gasparri is with the Department of Engineering, University of ÒRoma TreÓ, Via della Vasca Navale, 79. Roma, 00146, Italy (gasparri@dia.uniroma3.it).}}

\IEEEaftertitletext{\vspace*{-\baselineskip}}

\markboth{Submitted to the IEEE International Conference on Intelligent Robots and Systems (IROS) 2014}{Williams et.\ al.:  Route Swarm:  Wireless Network Optimization through Mobility}

\maketitle

\begin{abstract}
In this paper, we demonstrate a novel hybrid architecture for coordinating networked robots in sensing and information routing applications. The proposed \textit{IN}formation and \textit{S}ensing driven \textit{P}hys\textit{I}cally \textit{RE}configurable robotic network (INSPIRE), consists of a Physical Control Plane (PCP) which commands agent position, and an Information Control Plane (ICP) which regulates information flow towards communication/sensing objectives.  We describe an instantiation where a mobile robotic network is dynamically reconfigured to ensure high quality routes between static wireless nodes, which act as source/destination pairs for information flow.  The ICP commands the robots towards evenly distributed inter-flow allocations, with intra-flow configurations that maximize route quality.  The PCP then guides the robots via potential-based control to reconfigure according to ICP commands.  This formulation, deemed Route Swarm, decouples information flow and physical control, generating a feedback between routing and sensing needs and robotic configuration.  We demonstrate our propositions through simulation under a realistic wireless network regime.
\end{abstract}

\section{Introduction}
\IEEEPARstart{T}{raditional} work on distributed cooperation in robotics has focused on position and motion configuration of collections of robots using localized algorithms, typically involving iterative exchange of state variables with single-hop neighbors, e.g., research on swarming, flocking,  and formation control \cite{Gazi:2004, Olfati-Saber:2006, Fax:2004}.  The current state of the art on distributed cooperation in robotics, focused on using only localized communication, can effectively solve problems in scenarios where there are relatively simple global application related objectives that do not change over time. However, due to the difficulties in translating multiple dynamically varying global objectives into local control actions, the problem of utilizing these algorithms in more complex sensing and communication networks remains an open question.

A recent advance in distributed cooperation techniques offers promise in utilizing simple swarm-like mobility in coordinating more complex tasks.  A typical problem when considering only local communications is that global connectivity might be lost.  Recent research has shown that this global property can be recovered even through local interactions~\cite{Zavlanos:2008, Williams:2013uw, Yang:2010, Gasparri:TRO:2013}. We believe that this recent advance, enabling global connectivity to be maintained at all times while a collection of robots is moving, provides fundamental new opportunities as complex tasks can be decomposed into simple sub components, while maintaining overall network connectivity. 

Toward such goals, we first introduce at a high level a novel hybrid architecture for command, control, and coordination of networked robots for sensing and information routing applications, called INSPIRE (for \textbf{IN}formation and \textbf{S}ensing driven \textbf{P}hys\textbf{I}cally \textbf{RE}configurable robotic network). In the INSPIRE architecture, we propose two levels of control.  At the low level there is a Physical Control Plane (PCP), and at the higher level is an Information Control Plane (ICP). At the PCP, iterative local communications between neighboring robots is used to shape the physical network topology by manipulating agent position through motion.  At the ICP, more sophisticated multi-hop network algorithms enable efficient sensing and information routing (e.g., shortest cost routing computation, time slot allocation for sensor data collection, task allocation, clock synchronization, network localization, etc.). Unlike traditional approaches to distributed robotics, the introduction of the ICP provides the benefit of being able to scalably configure the sensing tasks and information flows in the network in a globally coherent manner even in a highly dynamical context by using multi-hop communications.

As a proof of concept of the INSPIRE architecture, we detail a simple instantiation, in which the robotic network is dynamically reconfigured in order to ensure high quality routes between a set of static wireless nodes (i.e.\ a \emph{flow}) while preserving connectivity, where the number and composition of information flows in the network may change over time.  In solving this problem, we propose ICP and PCP components that couple connectivity-preserving robot-to-flow allocations, with communication optimizing positioning through distributed mobility control; a heuristic we call \emph{Route Swarm}.  Finally, we demonstrate our propositions through simulation, illustrating the INSPIRE architecture and the Route Swarm heuristic in a realistic wireless network regime.

\section{State of the Art} \label{sec:art}

Distributed mobility control has been well investigated in the robotics community in recent years. In the context of multi-robot systems, distributed coordination protocols endow agents with simple local interactions, yet yield fundamentally useful collective behaviors.  Coordination algorithms can broadly be classified in three families, that is swarming, flocking and formation control. Swarming aims at achieving an aggregation of the team through local simple interaction~\cite{Gazi:2004},  flocking is a form of collective behavior of a large number of agents with an agreement in the direction of motion and velocity~\cite{Olfati-Saber:2006}, while formation control dictates the team reach a desired formation shape~\cite{Fax:2004}. For all of these objectives, potential-based control techniques represent an effective solution~\cite{Dimarogonas:2008}, combining provable performance with ease of control.  As robots are usually required to either communicate or sense each other for all time, the connectivity maintenance of the network topology also needs to be addressed.  Recent algorithms have been proposed to preserve the connectivity of the network topology over time, with approaches ranging from the control of addition and removal of edges~\cite{Williams:2013bh}, to the estimation and control of the algebraic connectivity~\cite{Gasparri:TRO:2013}.

The integration of mobile robotics and wireless networking is an emerging domain. Researchers have previously investigated deploying mobile nodes to provide sensor coverage in wireless sensor networks~\cite{Berman:TMC:2007, Gasparri:2008}.  In~\cite{Zavlanos:2008}, the authors present a work to ensure connectivity of a wireless network of mobile robots while reconfiguring it towards generic secondary objectives. Going beyond connectivity, recently, research has also addressed how to control a team of robots to maintain certain desired end-to-end rates while moving robots to do other tasks, referred to as the problem of maintaining network integrity~\cite{Zavlanos:ACC:2013}.  This is done by interleaving potential-field based motion control and at the higher level an iterative primal-dual algorithm for rate optimization.  All of these works point to the need for a hybrid control framework where low-level motion control can be integrated with a higher-level network control plane such as the INSPIRE architecture illustrated in this work\footnote{A complete characterization and general analysis of the INSPIRE architecture is the topic of our future work.}.
 
Closely related to our work is an early paper that advocated motion control as a network primitive in optimizing network information flows~\cite{Goldenberg04}.  Although related in spirit, we provide in this work fundamental advances in flow-to-flow reallocations, dynamic and flexible connectivity maintenance allowing network reconfigurability, and refined potential-based control that requires only inter-agent distance in optimizing intra-flow positioning.  Another, more recent work ~\cite{Mostofi12}, focuses on a single-flow setting, but considers a more detailed fading model communication environment, and a slightly different path metric.  In contrast to~\cite{Mostofi12}, we make novel contributions in multi-flow optimization which we have shown requires a more sophisticated network-layer information control plane.  Moreover, the motion control presented in~\cite{Mostofi12} can also be integrated with and adopted as a component of the PCP in the INSPIRE architecture presented here.  

\section{Background Material} \label{sec:pre}

To begin, we give an overview of the background material and assumptions necessary for our contributions in this work.

\subsection{Agent and Interaction Models}

\begin{figure}[t]
\centering
\includegraphics[width=2.5in]{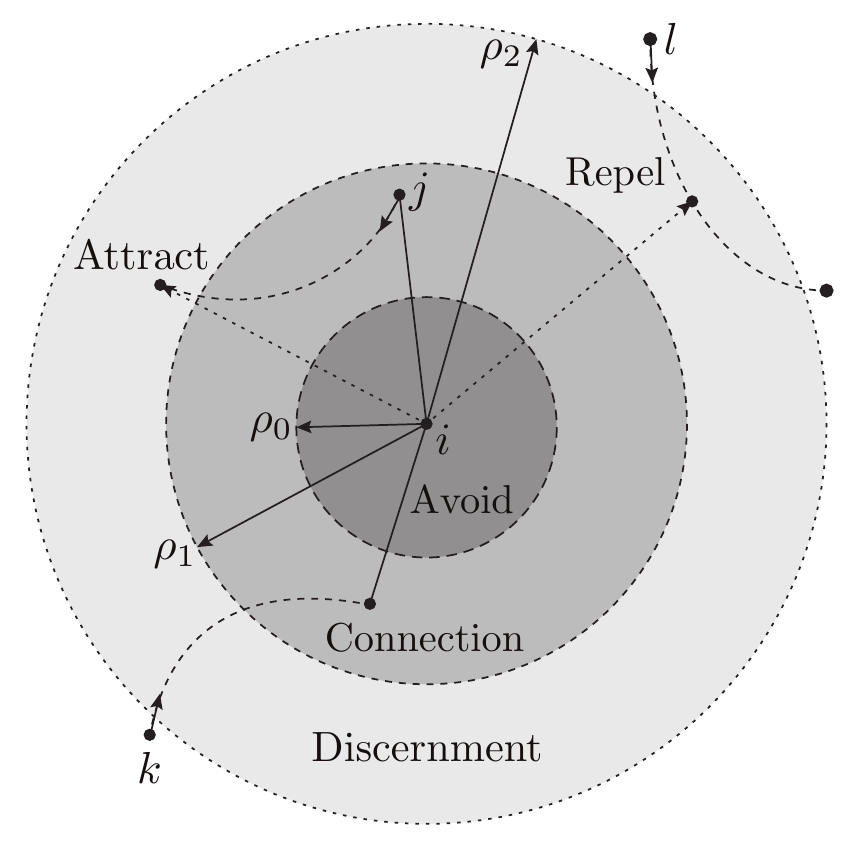}
\caption{Agent interaction model with radii determining sensing and communication $\norm{x_{ij}} \leq \rho_2$, neighbor decisions relative to constraints $\rho_1 < \norm{x_{il}} \leq \rho_2$, link establishment $\norm{x_{ik}} \leq \rho_1$, and collision avoidance $\norm{x_{ij}} \leq \rho_0$.}
\label{Fig-InteractRegions}
\end{figure}

Consider a system of $n = m+s$ agents consisting of $m$ mobile robots indexed by $\mc{I}_{M} \triangleq \{1,\ldots, m\}$, and $s$ static sensors indexed by $\mc{I}_{S} \triangleq \{m+1,\ldots, m+s\}$.  The mobile robots are assumed to have single integrator dynamics
\eqb{EQ-AgDyn}
\dot{x}_i = u_i
\eqe
where $x_i, u_i \in \mbb{R}^2$ are the position and the velocity control input for an agent $i \in \mc{I}_M$, respectively.  Assume that all agents can intercommunicate in a proximity-limited way, inducing interactions (or topology) of a time varying nature.  Specifically, letting $d_{ij} \triangleq \norm{x_{ij}} \triangleq \norm{x_i-x_j}$ denote the distance between agents $i$ and $j$, and $(i,j)$ a link between connected agents, the spatial neighborhood of each agent is partitioned by defining concentric radii $\rho_2 > \rho_1 > \rho_0$ as in \figurename \ref{Fig-InteractRegions}, where we refer to $\rho_2,\rho_1,\rho_0$ as the \emph{interaction}, \emph{connection}, and \emph{collision avoidance} radii, respectively.  The radii introduce a \emph{hysteresis} in interaction by assuming that links $(i,j)$ are established only after $d_{ij} \leq \rho_1$, with link loss then occurring when $d_{ij} > \rho_2$, generating the annulus of $\rho_2-\rho_1$ where \emph{decisions} on link additions and deletions are made (c.f.\ Section \ref{sec:cai}).

The above spatial interaction model is formalized by the \emph{undirected dynamic graph}, $\mbb{G} = (\mc{V},\mc{E})$, with vertices (nodes) $\mc{V}$ indexed by $\mc{I}_M \cup \mc{I}_S$ (the agents), and edges $\mc{E} \subseteq \mc{V} \times \mc{V}$ such that $(i,j) \in \mc{E} \Leftrightarrow (\norm{x_{ij}} \leq \rho_2) \; \wedge \; \sigma_{ij}$, with switching signals \cite{Ji:2007hu}:
\eqb{EQ-EdgeSwitch}
\sigma_{ij}= \left\{\begin{array}{ll}
0, & (i,j) \notin \mc{E} \; \wedge \; \norm{x_{ij}} > \rho_{1} \\
1, & \text{otherwise}
\end{array}\right.
\eqe
where $(i,i) \notin \mc{E}$ (no self-loops) and $(i,j) \in \mc{E} \Leftrightarrow (j,i) \in \mc{E}$ (symmetry) hold for all $i,j \in \mc{V}$.  Nodes with $(i,j) \in \mc{E}$ are called \emph{neighbors} and the neighbor set for an agent $i$ is denoted $\mc{N}_i = \{j \in \mc{V} \st (i,j) \in \mc{E}\}$.

\subsection{Assumptions and Problem Formulation}

From the set of static sensors $\mc{I}_S$ we construct $f$ \emph{information flows} indexed by $\mc{I}_{F} \triangleq \{n+1, \ldots, n+f\}$, each consisting of source-destination pairs defining a desired flow of network information.  For a given flow $i \in \mc{I}_F$, we use the following notation:  $\mc{F}_i \in \mc{I}_{S} \times \mc{I}_{S}$ represents the source and destination nodes for flow $i$, with $\mc{F}_i^s \in \mc{I}_{S}$ and $\mc{F}_i^d \in \mc{I}_{S}$ representing source and destination indices, respectively.  Further, for convenience we use notation $x_i^s, x_i^d \in \mbb{R}^2$ to represent the position of the source and destination for the flow $i \in \mc{I}_{F}$.  The set of flow pairs is denoted $\mc{F} \triangleq \{\mc{F}_1, \ldots, \mc{F}_f\}$.  At any time, a subset of these static pairs is active, forming the set of active flows $\mc{I}_{F}$, calling for dynamic configurability of the hybrid network, our contribution in this work.  Thus, at a high level our system objective is to facilitate information flow for each source/destination pair by configuring the mobile robots such that each flow is \emph{connected} and is at least approximately optimal in terms of data transmission, and that the entire network itself $\mbb{G}$ is connected to guarantee complete network collaboration.

\begin{assumption}[Connectedness]\label{as:connect}
It is assumed that the locations of the static nodes and the cardinality of mobile nodes is such that the set of connected graphs $\mbb{G}$ that could be formed is non-empty, and that their communication graph is initiated to be in this set.
\end{assumption}

To measure link quality towards optimizing a given flow, we assume that each $(i,j) \in \mc{E}$ has a weight parameter $w_{ij}$ that describes their cost with respect to transmitting information.  A commonly used metric for link quality is \emph{ETX}, i.e.\ the expected number of transmissions per successfully delivered packet.  This can be modeled as the inverse of the successful packet reception rate $\lambda_{ij}$ over the link.  As the expected packet reception rate has been empirically observed and analytically shown to be a sigmoidal function of distance decaying from 1 to 0 as distance $d_{ij}$ is increased~\cite{ZunigaTOSN}, it can be modeled as follows:
\eqb{eq:recept}
\lambda_{ij} \approx 1 - \frac{1}{(1 + e^{-a (d_{ij} - b)})}
\eqe
where $a,b \in \mbb{R}_{+}$ are shape and center parameters depending on the communication range and the variance of environmental fading.  Accordingly, the link weights $w_{ij}$, if chosen to represent ETX, can be modeled as a convex function of the inter-node distance $d_{ij}$:
\eqb{eq:linkweight}
w_{ij} = \frac{1}{\lambda_{ij}} = \frac{1}{1 - \frac{1}{1 + e^{- a (d_{ij} - b)}}} = 1 + e^{a (d_{ij} - b)}
\eqe

The cost for flow $k \in \mc{I}_F$ is then taken to be the sum of ETX values on the path of the flow, i.e.:
\eqb{eq:flowweight}
W_k = \sum_{(i,j) \in \mc{E}_F^k} w_{ij}
\eqe
where we apply notation $\mbb{G}_F^k= (\mc{V}_F^k, \mc{E}_F^k)$ as the graph defining the interconnection over flow $k \in \mc{I}_F$ (we give a concrete definition of flow \emph{membership} in Section \ref{sec:routeswarm}).  Our problem in this work is then formalized as follows:

\begin{problem}[Multi-flow optimization]\label{prob:flowopt}
The network-wide goal then is to find an allocation and configuration of mobile agents so as to minimize the total cost function\footnote{The negative of the cost could be treated as a utility function. We therefore equivalently talk about cost minimization or utility maximization.} $\sum_{k =1}^f W_k$ while maintaining both intra-flow connectivity (to guarantee information delivery) and inter-flow connectivity (to guarantee flow-to-flow information passage/collaboration).
\end{problem}

\section{Information Control Plane (ICP)} \label{sec:icp}

There are two key elements in solving Problem \ref{prob:flowopt}: on the one hand, within a given flow $k \in \mc{I}_F$, for a given allocation of a certain number of mobile nodes to that flow, node configuration should minimize the flow cost $W_k$. On the other hand, the number of mobile nodes allocated to each flow should minimize the overall cost $\sum_k W_k$.  We first consider these optimizations ideally, in the absence of connectivity constraints. The first, per-flow element of the network optimization dictates the desired spatial configuration of allocated mobile nodes within a flow.

\begin{theorem}[Equidistant optima] \label{thrm:equi}
For a fixed number of mobile nodes $m_k \triangleq \abs{\mc{V}_F^k}$ allocated to a given flow and arranged on the line between the source and destination of that flow, the arrangement which minimizes \eqref{eq:flowweight} is one where the nodes are equally spaced.
\end{theorem}

\begin{IEEEproof}
This follows from the following general result: to minimize a convex $y(\overrightarrow{z}$) s.t. $\sum{z_i} = c$, the first order condition (setting the partial derivative of the Lagrangian with respect to each element $z_i$ to 0)  yields that $\frac{\partial y}{z_i} = \mu$ where $\mu$ is the Lagrange multiplier. Now if $\frac{\partial y}{\partial z_i}$ is the same for all $z_i$, then the solution to this optimization is to set $z_i = \frac{c}{\abs{z}}$, i.e.\ all variables are made to be the same.

The $z_i$ in the above correspond to the inter-node distances $d_{ij}$, and the $y$ corresponds to \eqref{eq:flowweight}. Since $w_{ij}$ is a convex function of distance between neighboring nodes, the path metric $W_k$ is a convex function of the vector of inter-node distances. Further, since the sum of all inter-node distances is the total distance between the source and destination of the flow, which are static, it is constrained to be a constant $d_k$.  Since the weight of each link is the same function of the inter-node distance, $\frac{\partial W_k}{d_{ij}}$ is the same for all pairs of neighboring nodes. Therefore the intra-flow optimization (i.e., choosing node positions to minimize $W_k$)  is achieved by a equal spacing of the nodes.
\end{IEEEproof}

The second, global cross-flow element of the network optimization dictates the number of mobile nodes allocated to each flow. The goal is to minimize the total network cost $\sum_k W_k$. Our approach to solving this optimization is motivated by the following observation: When the intra-flow locations of the robots are optimized to be equally spaced, $W_k$ is a function of the number of nodes $m_k$ allocated to flow $k$, and the total number of nodes allocated to all flows is constrained by the total number of mobile nodes. If we could show that $W_k$ is a convex function, then to minimize the total network cost we need to identify the allocation at which the marginal costs for all flows are as close to equal as possible (ideally, if $m_k$ was a continuous quantity, they would all be equal at the optimum point, but due to the discrete nature of $m_k$ this is generally not possibly).

In the following, for ease of analysis, we consider the continuous relaxation of the problem, allowing $m_k$ to be a real number, and hence $W_k$ to be a continuous function. 

\begin{theorem}[Convexity] \label{thrm:convex}
For any flow $k$ that has been optimized to have the lowest possible cost $W_k$ (i.e. all $m_k$ mobile nodes are equally spaced), $W_k$ is a convex function of $m_k$.
\end{theorem}

\begin{IEEEproof}
\begin{eqnarray}
W_k &=& (m_k+1) w \lp\frac{d_k}{m_k +1}\rp\\
\Rightarrow W_k' & = & w \lp \frac{d_k}{m_k + 1} \rp -  \frac{d_k}{m_k+1}w'\lp\frac{d_k}{m_k + 1} \rp\\
\Rightarrow W_k'' & = &  w '' \lp\frac{d_k}{m_k + 1}\rp \frac{d^2}{(m_k+1)^3}
\end{eqnarray}
Since the link weight function $w(\cdot)$ is a convex function of the internode distance, we have that $w''(\frac{d_k}{m_k + 1})$ is positive, and therefore we have that $W_k''$ is positive, hence $W_k$ is convex in $m_k$.
\end{IEEEproof}

Note that in reality $m_k$ is a discrete quantity, but the above argument suffices to show that $W_k$ is a discrete convex function of $m_k$ (since convexity over a single real variable implies discrete convexity over the integer discretization of the variable).

\begin{theorem}[Optimality] \label{thrm:optimal}
The problem of minimizing $\sum_k W_k(m_k)$ subject to a constraint on $\sum_k m_k$ can be solved optimally in an iterative fashion by the following greedy algorithm: at each iteration, move one node from the flow where the removal induces the lowest increase in cost to the flow where its addition would yield the highest decrease in cost, so long as the latter's decrease in cost is strictly higher in absolute value than the former's increase in cost (i.e. so long as the move serves to reduce the overall cost).
\end{theorem}

We omit the detailed proof due to space constraint, but intuitively, this algorithm works by moving the system iteratively towards the optimum by following the steepest gradient in terms of cost reduction for the movement of each node. Since the overall optimization problem is convex, there is only a single optimum, to which this algorithm will converge. Moreover, since there is a strict improvement in each step and there are a finite number of nodes, the algorithm reaches the optimal arrangement in a finite number of steps.

Thus far, we have described both the intra-flow and inter-flow optimization problems in an ideal setting where both problems are convex optimization problems and as such can be solved exactly. However, in the robotic system we are considering there is one significant source of non-ideality/non-convexity, which is that the network must be maintained at all times in a connected configuration. This has two consequences. First, some of the mobile nodes may be needed as \emph{bridge} nodes that do not participate in any flow and are instead used to maintain connectivity across flows. Second, the locations of some of the nodes even within each flow may be constrained in order to maintain the connectivity requirement. The solution for the constrained problem therefore may not correspond exactly to the solutions of the ideal optimization problems described above.

We therefore develop a heuristic solution that we refer to as \emph{Route Swarm}, which is inspired by and approximates the ideal optimizations above but is adapted to maintain connectivity. Route Swarm has both an intra-flow and inter-flow component. As per the INSPIRE architecture, the intra-flow function is performed by the PCP, while the inter-flow function is performed by ICP.  The ICP algorithm, shown as pseudocode as Algorithm~\ref{Alg-ICP}, approximates the ideal iterative optimization described above by allocating mobile robots between flows greedily on the basis of greatest cost-reduction; to handle the inter-flow connectivity constraint, it incorporates a subroutine (Algorithm~\ref{Alg-DetectBridges}) to detect which nodes are not mobile because they must act as bridge nodes. Moreover, it also allocates any nodes that are no longer required to support an inactive flow to join active flows.  And within each flow, the PCP algorithm (Algorithm~\ref{alg:pcp}) attempts to keep the robots as close to evenly spaced as possible while taking into account the inflexibility of the bridge nodes.  The details of the route swarm algorithm are given below.

\subsection{The Route Swarm Heuristic} \label{sec:routeswarm}

\begin{algorithm}[t]

\begin{algorithmic}[1]
\Procedure{InformationControlPlane}{}
	\State $\triangleright$ Detect initial flow members based on shortest paths:
	\For{$i \in \mc{I}_F$}
		\State $\mc{M} \gets $ \Call{ShortestPath}{$\mbb{G}, \mc{F}_i$}
	\EndFor
	\State $\triangleright$ Detect flow-to-flow bridges for connectivity:
	\State $b \gets$ \Call{DetectBridges}{$\mbb{G}, \mc{F}, \mc{M}$}
	\State $\triangleright$ Detect best connectivity-preserving flow detachment:
	\State $d \gets$ \Call{BestDetachment}{$\mbb{G}, \mc{F}, \mc{M}, b$}
	\State $\triangleright$ Compute flow attachment with most utility:
	\State $a \gets$ \Call{BestAttachment}{$\mc{F}, \mc{M}$}
	\State $\triangleright$ Ensure optimizing reconfiguration exists (weighted by $\beta \in \mbb{R}_{+})$:
	\If{$a > \beta d$}
		\State $\triangleright$ Optimal command is best detach/attach pair:
		\State \Return $\mc{C}_d \gets a$
	\EndIf
\EndProcedure
\end{algorithmic}
\caption{ICP optimization algorithm.}
\label{Alg-ICP}
\end{algorithm}

\begin{algorithm}[t]

\begin{algorithmic}[1]
\Procedure{DetectBridges}{$\mbb{G}, \mc{F}, \mc{M}$}
	\State $\triangleright$ Initialize supergraph with nodes for each flow:
	\State $\mbb{S} \gets (\{i \in \mc{I}_F\}, \emptyset)$
	\State $\triangleright$ Append nodes/edges for non-flow members:
	\State $\mbb{S} \gets$ \Call{AddNonFlowMembers}{$\mbb{S}, \mc{F}, \mc{M}$}
	\State $\triangleright$ Append edges for flow members:
	\State $\mbb{S} \gets$ \Call{AddFlowMembers}{$\mbb{S}, \mc{F}, \mc{M}$}
	\State $\triangleright$ Bridges lie on shortest path between all flow pairs:
	\State \Return $b \gets$ \Call{ShortestPaths}{$\mbb{S}, \mc{F}$}
\EndProcedure
\end{algorithmic}
\caption{ICP bridge detection.}
\label{Alg-DetectBridges}
\end{algorithm}

\begin{figure}[t]
\centering
\includegraphics[width=3.35in]{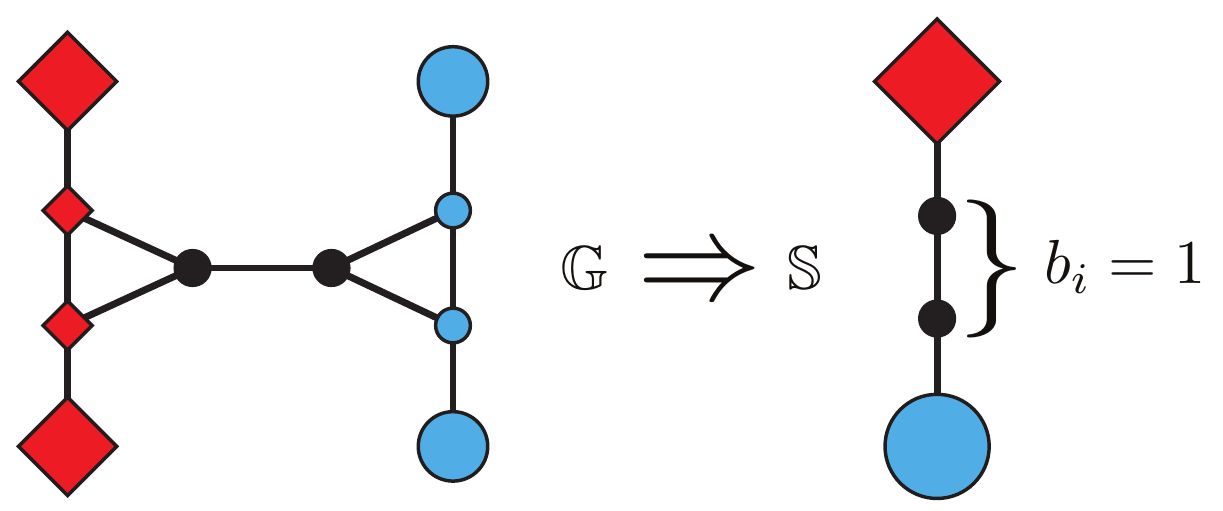}
\caption{Example supergraph construction for a network with $f = 2$ flows, $s = 4$ static nodes, and $m = 6$ mobile robots.  Flow membership is denoted by color and node shape.  Notice that a multi-hop bridge is detected via the shortest path between flows.}
\label{Fig-SupergraphExample}
\end{figure}

In solving the connectivity constrained version of the flow optimization of Theorems \ref{thrm:equi}, \ref{thrm:convex}, and \ref{thrm:optimal}, due to problem complexity we provide a heuristic algorithm to solve the inter-flow allocation problem, leaving the intra-flow optimization to the PCP (Section \ref{sec:pcp}).  Algorithm~\ref{Alg-ICP} depicts the high-level components of the proposed heuristic, each of which is detailed in the sequel.

To begin, we define   the flow membership for an agent as $i \in \mc{I}_{M}$ as $\mc{M}_i \in \mc{I}_{F}$, where the set of memberships is denoted as $\mc{M} \triangleq \{\mc{M}_1, \ldots, \mc{M}_n\}$. We denote with $\mbb{G}_F^i = (\mc{V}_F^i, \mc{E}_F^i)$ the graph defining the interconnection over flow $i \in \mc{I}_F$ of agents $j \in \mc{I}_S \cup \mc{I}_M$ with $i \in \mc{M}_j$.  Thus we have $\mc{V}_F^i = \{j \in \mc{I}_S \cup \mc{I}_M \st i \in \mc{M}_j\}$ and $(j,k) \in \mc{E}_F^i \Leftrightarrow (i \in \mc{M}_j \cap \mc{M}_k) \wedge (j \in \mc{N}_k)$, with $k \in \mc{I}_S \cup \mc{I}_M$.  Notice that by definition $\mbb{G}_F^i \subseteq \mbb{G}$.  We will also refer to the collection of flow graphs simply by $\mbb{G}_F$.  We also have flow neighbors defined as $\mc{N}_i^F = \{j \in \mc{N}_i \st \mc{M}_j \cap \mc{M}_i \neq \emptyset\}$. Furthermore, we denote with $\mc{D}_i \in \mc{I}_M$  the detachable agents $j \in \mc{I}_M$ for flow $i \in \mc{I}_F$, i.e.\ those agents for which reconfiguration does not impact network connectedness.  The set of detachments is denoted by $\mc{D} \triangleq \{\mc{D}_1, \ldots, \mc{D}_f\}$.

Due to the system connectivity constraint, the ICP is not free to select \emph{any} mobile agent for flow reallocation, specifically as intra-flow connectivity or overall flow-to-flow connectivity may be lost.  Thus, the ICP must detect \emph{bridges}, i.e., mobile nodes whose reconfiguration might break flow-to-flow connectivity over the network, and also consider \emph{safe detachments} (flow-to-flow motion), i.e., a node whose reconfiguration in the workspace does not impact the connectivity of the source flow.

Respecting our connectivity constraint, we first detect initial flow membership by computing the shortest path (in terms of link cost) for each source/destination pair per flow by using for example Dijkstra's algorithm (lines 3-5 of Algorithm \ref{Alg-ICP}). This defines the connected \emph{backbone} for each flow implicitly identifying the mobile nodes required to maintain flow-connectivity.  Additionally, we have optimality of the connected flow backbones as we maximize the link utility (or minimize path costs) between source and destination nodes.

In detecting bridge agents we follow the process outlined by Algorithms~\ref{Alg-DetectBridges},~\ref{Alg-AddNonFlowMembers}, and~\ref{Alg-AddFlowMembers}, where we denote by $b_i \in \{0,1\}$ the status of agent $i \in \mc{I}_S \cup \mc{I}_M$ as a connectivity preserving bridge, with $b = \{b_1, \ldots, b_n\}$. Briefly speaking, the primary problem of this process is the construction of a \emph{supergraph}, denoted $\mbb{S} = (\mc{V}_{\mbb{S}}, \mc{E}_{\mbb{S}})$, defining the interconnection of the flows.  In this way, we can identify nodes which are critical in defining the flow-to-flow connectivity over the system.  Figure~\ref{Fig-SupergraphExample} depicts an example of supergraph construction. We first add one node for each flow in the system, and then we add a node for each non-flow member in the system that represents potential bridge candidates. Edges between non-flow members are preserved, while there is only one edge between any given non-flow member and the node which represents a flow in the supergraph.  Bridges can then be simply detected as the members of the shortest path between any pair of nodes representing flows in the supergraph (e.g.\ \figurename \ref{Fig-SupergraphExample}), where connectivity is guaranteed by construction:

\begin{proposition}[Bridge detection] \label{prop:bridge}
Consider the graph $\mbb{G}_c \subseteq \mathbb{G}$ obtained by including all the flow-members for any flow and the non-flow members belonging to any shortest path of the supergraph $\mathbb{S}$. Then $\mbb{G}_c$ is a spanning-graph representing a connected component of the graph $\mbb{G}$.
\end{proposition}

\begin{IEEEproof}
In order to prove this result we must show that both intra-flow and inter-flow connectedness is ensured. The former follows directly from  the flow-membership definition  while the latter follows from the connectedness between any pair of flows.
\end{IEEEproof}

After identifying the bridge agents that maintain flow-to-flow connectivity and the safely detachable agents per-flow (those which are not on the backbone), the ICP issues reconfiguration commands $\mc{C}_i \in \mc{I}_{F}$ to mobile agents $i \in \mc{I}_M$ indicating a desired flow membership towards optimizing inter-flow agent allocation.  Specifically, as detailed in Algorithms \ref{Alg-BestDetachment} and \ref{Alg-BestAttachment}, and as motivated by Theorem \ref{thrm:optimal}, we compute the safe detachment having the least in-flow utility, and couple it with the flow attachment (i.e.\ the addition of a contributing flow member) which improves most in terms of link utility and flow alignment (the primary contributions a mobile agent can have in information flow).  This decision, if feasible (utility of attachment outweighs cost of detachment), is then passed to the PCP to execute the mobility necessary for reconfiguration, achieving our goal of dynamic utility improving and connectivity preserving network configurations.

\begin{algorithm}[t]

\begin{algorithmic}[1]
\Procedure{AddNonFlowMembers}{$\mbb{S}, \mc{F}, \mc{M}$}
	\State $\triangleright$ Add nodes for each non-leaf, non-flow member:
	\State $\mc{V}_{+} \gets \{i \in \mc{I}_M \st (\mc{M}_i = \emptyset) \wedge (\mc{N}_i \geq 2)\}$
	\State $\mc{V}_{\mbb{S}} \gets \mc{V}_{\mbb{S}} \cup \mc{V}_{+}$
	\State $\triangleright$ Add edges between non-flow members:
	\State $\mc{E}_{\mbb{S}} \gets \mc{E}_{\mbb{S}} \cup \{(i,j) \st (i,j \in \mc{V}_{+}) \wedge (j \in \mc{N}_i)\}$
	\State $\triangleright$ Add non-flow member to flow member edges:
	\State $\mc{E}_{\mbb{S}} \gets \mc{E}_{\mbb{S}} \cup \{(i,j) \st (i \in \mc{V}_{+}) \wedge (j \in \mc{N}_i) \wedge (\mc{M}_j \neq \emptyset)\}$
	\State \Return $\mbb{S}$
\EndProcedure
\end{algorithmic}
\caption{ICP supergraph non-flow member nodes/edges.}
\label{Alg-AddNonFlowMembers}
\end{algorithm}

\begin{algorithm}[t]

\begin{algorithmic}[1]
\Procedure{AddFlowMembers}{$\mbb{S}, \mc{F}, \mc{M}$}
	\State $\triangleright$ Add edges due to multiple flow memberships:
	\For{$i \in \mc{I}_M \st \abs{\mc{M}_i} \geq 2$}
		\State $\triangleright$ Add an edge for every membership pair:
		\State $\mc{E}_{\mbb{S}} \gets \mc{E}_{\mbb{S}} \cup \{(i,j) \st j \in \mc{I}_F \cap \mc{M}_i\}$
	\EndFor
	\State \Return $\mbb{S}$
\EndProcedure
\end{algorithmic}
\caption{ICP supergraph flow member edges.}
\label{Alg-AddFlowMembers}
\end{algorithm}

\begin{algorithm}[t]

\begin{algorithmic}[1]
\Procedure{BestDetachment}{$\mbb{G}, \mc{F}, \mc{M}, b$}
	\State $\triangleright$ Non-member, non-bridges are detachable:
	\For{$(i \in \mc{I}_M) \wedge (\mc{M}_i = \emptyset) \wedge (\lnot b_i)$}
		\State $j \gets \; \text{Flow most contributed to by} \;\text{agent}\; i \;\text{(utility)}$
		\State $\mc{M}_i \gets j$
		\State $\mc{D}_j \gets \mc{D}_j \cup i$
	\EndFor
	\State $\triangleright$ Best detachment has least in-flow utility:
	\State \Return $\text{argmin}_{i \in \mc{D}}(\sum_{j \in \mc{N}_i^F}w_{ij})$
\EndProcedure
\end{algorithmic}
\caption{ICP best flow detachment.}
\label{Alg-BestDetachment}
\end{algorithm}

\begin{algorithm}[t]

\begin{algorithmic}[1]
\Procedure{BestAttachment}{$\mc{F}, \mc{M}$}
	\State $\triangleright$ Determine utility of attachment per flow:
	\For{$i \in \mc{I}_F$}
		\State $\triangleright$ Weigh added node utility against flow path alignment:
		\State $a_i \gets (\abs{\mc{V}_F^i}+1)\sum_{j \in \mc{V}_F^i}1/\norm{x_j-1/2(x_i^s+x_i^d)}$
	\EndFor
	\State \Return $\text{argmax}_{i \in \mc{I}_F}(a_i)$
\EndProcedure
\end{algorithmic}
\caption{ICP best flow attachment.}
\label{Alg-BestAttachment}
\end{algorithm}

\section{Physical Control Plane (PCP)} \label{sec:pcp}

The complementary component to the ICP in the INSPIRE architecture is the PCP which coordinates via state feedback, i.e.\ $\{\mb{x}, \mbb{G}, \mbb{G}_F\}$, to generate swarming behaviors that optimize the network dynamically in response to ICP commands.  Our desire for generality in coordinating behaviors dictates that the PCP takes on a \emph{switching} nature, associating a distinct \emph{behavior controller} with each of a finite set of discrete agent states.  Specifically,  define
\eqb{EQ-PCPStates}
\mc{S}_i \in \mbb{B} \triangleq \{\textit{SWARMING}, \textit{RECONFIGURE}\}
\eqe
as the \emph{behavior state} of a mobile robot $i \in \mc{I}_{M}$, where $\mbb{B}$ is the space of discrete agent behaviors.  Here, when $\mc{S}_i = \textit{SWARMING}$, agent $i$ acts to optimize its assigned flow $\mc{M}_i$.  Otherwise, when $\mc{S}_i = \textit{RECONFIGURE}$, agent $i$ traverses the workspace fulfilling global allocation commands $\mc{C}_i$ from the ICP.   In this work, the state machine that drives the switching of the behavior controllers is depicted in Algorithm \ref{alg:pcp}.  Each component comprising the PCP switching is detailed in the sequel.

\begin{algorithm}[t]

\begin{algorithmic}[1]
\Procedure{PhysicalControlPlane}{}
	\For{$i \in \mc{I}_M$} \Comment{Control mobile agents}
	\State $\triangleright$ Reconfiguration Commanded:
		\If{$\mc{C}_i \neq \emptyset$}
			\State Set waypoint to target flow $\mc{C}_i \in \mc{I}_F$
			\State $\mc{S}_i \gets \textit{RECONFIGURE}$
		\EndIf
		\State $\triangleright$ Flow members, to-flow maneuver:
		\If{$\mc{M}_i \neq \emptyset \wedge \mc{S}_i \neq \textit{RECONFIGURE}$}
			\If{$\lnot$(\textit{On Path Connecting Flow} $\mc{M}_i$)}
				\State Set waypoint to flow $\mc{M}_i \in \mc{I}_F$
				\State $\mc{S}_i \gets \textit{RECONFIGURE}$
			\Else
				\State $\mc{S}_i \gets \textit{SWARMING}$
			\EndIf
		\EndIf
		\State $\triangleright$ Agent Behaviors:
		\If{$\mc{S}_i = \textit{SWARMING}$}
			\State Run dispersion controller optimizing flow $\mc{M}_i$
		\EndIf
		\If{$\mc{S}_i = \textit{RECONFIGURE}$}
			\State Run waypoint controller for reconfiguration
			\If{\textit{At Waypoint}}
				\State $\mc{S}_i \gets \textit{SWARMING}$
			\EndIf
		\EndIf
	\EndFor
\EndProcedure
\end{algorithmic}
\caption{PCP switching logic.}
\label{alg:pcp}
\end{algorithm}

\subsection{Constraining Agent Interaction} \label{sec:cai}

In order to control the properties of $\mbb{G}$ (i.e.\ connectivity) we exploit the \emph{constrained interaction} framework proposed by Williams and Sukhatme in \cite{Williams:2013bh}.  The constrained interaction framework acts through hysteresis \eqref{EQ-EdgeSwitch} to regulate links \emph{spatially} with simple application of attraction and repulsion to retain established links or reject new links with respect to topological constraints.  Define the \emph{discernment region} $\norm{x_{ij}} \in (\rho_1,\rho_2]$, where agent $i$ decides relative to system constraints (here connectivity) whether agent $j$ is a \emph{candidate} for link addition ($j \notin \mc{N}_i$) or deletion ($j \in \mc{N}_i$), or if agent $j$ should be attracted (retain $(i,j) \in \mc{E}$) or repelled (deny $(i,j) \notin \mc{E}$).  Define \emph{predicates} for link addition and deletion, $P_{ij}^a, P_{ij}^d: \mc{V} \times \mc{V} \rightarrow \{0,1\}$, activated at $\rho_2$ and $\rho_1$, respectively, that indicate constraint violations if the link $(i,j)$ were allowed to be either created or destroyed, i.e.\ $\norm{x_{ij}}$ transits $\rho_1$ or $\rho_2$.  The predicates designate for the $i$th agent the membership of nearby agents in link addition and deletion candidate sets $\mc{C}_{i}^a, \mc{C}^d_{i}$, and attraction and repulsion sets $\mc{D}_{i}^a, \mc{D}_{i}^r$.  Link control is then achieved by choosing control $u_i$ having attractive and repulsive \emph{potential fields} between members of $\mc{D}_{i}^a, \mc{D}_{i}^r$, respectively.

In particular, to regulate network topology spatially, we design the agent controls as follows:
\eqb{EQ-ConIntControls}
u_i = u_i^e + u_i^o-\nabla_{x_{i}} \lp\sum_{j \in \mc{D}_i^a} \psi_{ij}^a + \sum_{j \in \mc{D}_i^r} \psi_{ij}^r + \sum_{j \in \Pi_i} \psi_{ij}^{\text{c}}\rp
\eqe
with \emph{potentials} $\psi_{ij}^a, \psi_{ij}^r, \psi_{ij}^c \colon \mbb{R}_+ \rightarrow \mbb{R}_+$, serving the purposes of attraction, repulsion, and collision avoidance, where $\Pi_i = \{j \in \mc{V} \st \norm{x_{ij}} \leq \rho_0\}$ is the collision avoidance set for agent $i$.  Further, each agent can also apply (based on $\mc{S}_i$) an exogenous objective (i.e.\ non-cooperative) controller $u_i^e \in \mbb{R}^2$ and an inter-agent coordination objective $u_i^o \in \mbb{R}^2$ (e.g.\ dispersion as will be seen in Section \ref{ss:intraflow}).  An appropriate attractive potential which we adopt for this work takes the following form:
\eqb{EQ-AttPot}
\psi_{ij}^a =  \frac{1}{\rho_2^{2} - \norm{x_{ij}}^{2}} + \Psi_a,  \quad \text{if} \quad \norm{x_{ij}} \in [\rho_1,\rho_2)
\eqe
where $\Psi_a(\norm{x_{ij}})$ is chosen such that \eqref{EQ-AttPot} is smooth over the $\rho_1$ transitions.  Similar to the attractive potential \eqref{EQ-AttPot}, the repulsive potential takes the form
\eqb{Eq-RepPot}
\psi_{ij}^r =
\displaystyle \frac{1}{\norm{x_{ij}}^{2}-\rho_1^{2}} + \Psi_r, \quad \text{if} \quad  \norm{x_{ij}} \in (\rho_1, \rho_2)
\eqe
where $\Psi_r$ is chosen to guarantee $\psi_{ij}^r$ is smooth over the $\rho_2$ transition.  Finally, a basic collision avoidance is given by potential
\eqb{EQ-CollAvoid}
\psi_{ij}^c =
\displaystyle \frac{1}{\norm{x_{ij}}^{2}} + \Psi_c, \quad \text{if} \quad \norm{x_{ij}} \in (0, \rho_0) \\
\eqe
with $\Psi_r$ chosen to guarantee $\psi_{ij}^c$ is smooth over the $\rho_0$ transition.

The attractive and repulsive potentials are constructed such that $\psi_{ij}^a \rightarrow \infty$ as $d_{ij} \rightarrow \rho_2$ and $\psi_{ij}^r \rightarrow \infty$ as $d_{ij} \rightarrow \rho_1$, guaranteeing link retention and denial, respectively, and allowing us through predicates $P_{ij}^a, P_{ij}^d$ to control desired properties of $\mbb{G}$ (e.g.\ connectivity).

\subsection{Connectivity Maintenance}

Notice that in \emph{maintaining} network connectivity, we require \emph{only}  link retention action, allowing us to immediately choose
\eqb{EQ-AddPred}
P_{ij}^a \triangleq 0, \quad \fa i \in \mc{I}_M, j \in \mc{I}_M \cup \mc{I}_S, (\mc{S}_i, \mc{S}_j) \in \mbb{B} \times \mbb{B}
\eqe
for the link addition predicates, effectively allowing link additions to occur between all interacting agents across all network states\footnote{Although in this work we allow all link additions, link addition control could be useful for example in regulating neighborhood sizes to mitigate spatial interference, or to disallow interaction between certain agents.}.

Now, in accordance with Algorithm \ref{alg:pcp}, the link deletion predicates are given by:
\eqb{EQ-DelPred}
\renewcommand*{\arraystretch}{1.5}
P_{ij}^d \triangleq \left\{\begin{array}{ll}
1, & (\mc{S}_i \vee \mc{S}_j = \textit{SWARMING}) \, \wedge \\
& ((\mc{M}_i \cap \mc{M}_j \neq \emptyset) \vee (b_i \vee b_j = 1)) \\
0, & \text{otherwise}
\end{array}\right.
\eqe
where by assumption $i,j \in \mc{I}_M$, i.e.\ only mobile agents apply controllers.  We choose link retention \eqref{EQ-DelPred} to guarantee that connectivity is maintained both within flows across $\mbb{G}_F$, and from flow to flow across bridge agents over the supergraph $\mbb{S}$, noting that idle agents with $\mc{M}_i = \emptyset$ and reconfiguring agents with $\mc{S}_i = \textit{RECONFIGURE}$ are free to lose links as they have been deemed redundant by the ICP with respect to network connectivity.

\subsection{Flow Reconfiguration Maneuvers}

While maintaining connectivity as above, each agent further acts according to ICP reconfiguration commands towards optimizing inter-flow allocations.  Specifically, in response to command $\mc{C}_i$, agent $i$ enters the reconfiguration state $\mc{S}_i \gets \textit{RECONFIGURE}$, and begins to apply a \emph{waypoint} controller as follows (c.f.\ lines 4-7 of Algorithm \ref{alg:pcp}).  When $\mc{S}_i = \textit{RECONFIGURE}$, agent $i$ applies exogenous objective controller
\eqb{EQ-WayControl}
u_i^e \triangleq \frac{x^w-x_i}{\norm{x^w-x_i}} - \dot{x}_i
\eqe
where $x^w \in \mbb{R}^2$ is the target waypoint calculated as the midpoint of target flow $\mc{C}_i$.

The input \eqref{EQ-WayControl} is a velocity damped waypoint seeking controller, having unique critical point $x_i \rightarrow x^w$ (i.e.\ a point at which $u_i^e = \mb{0}$), guaranteeing that the target intra-flow positioning (and thus membership) for agent $i$ is achieved.  As the convergence of $x_i \rightarrow x^w$ is asymptotic in nature, to guarantee finite convergence and state switching, we apply a saturation $\norm{x^w-x_i} \leq \epsilon^w$ with $0 < \epsilon^w << 1$ to detect waypoint convergence, initiating a switch to $\mc{S}_i \gets \textit{SWARMING}$ as in lines 23-25 of Algorithm \ref{alg:pcp}.

\subsection{Intra-Flow Controllers}\label{ss:intraflow}

Once the ICP has assigned flow memberships $\mc{M}_i \fa i \in \mc{I}_M$ and all \emph{commanded} reconfigurations $\mc{C}_i$ have been completed, the mobile agents begin to seek to optimize the flow to which they are a member.  First, we assume that flow members must configure along the line segment connecting flow source/destination pairs, yielding in the case of proximity-limited communication, a line-of-sight or beamforming style heuristic.  The membership of an agent $i \in \mc{I}_M$ to a flow $j \in \mc{M}_i$ thus initiates a check to determine if $x_i$ lies on the flow path $x_j^s + \tau x_j^d$, within a margin $0 < \epsilon_F << 1$ (c.f.\ lines 9-16, Algorithm \ref{alg:pcp}).  To do so, the projection of $x_i$ onto $x_j^s + \tau x_j^d$ is determined first by computing
\eqb{EQ-Project1}
\tau \triangleq \frac{(x_i-x_j^s)\cdot(x_j^d-x_j^s)}{\norm{x_j^d-x_j^s}^2}
\eqe
defining whether the projection will lie within or outside of the flow path.  Then we have the saturated projection
\eqb{EQ-FlowProject}
\renewcommand*{\arraystretch}{1.5}
x_{i\rightarrow \mc{F}_j} = \left\{\begin{array}{ll}
x_j^s-\alpha\tau(x_j^d-x_j^s), & \tau < 0\\
x_j^d-\alpha\tau(x_j^d-x_j^s), & \tau > 1\\
x_j^s+\tau(x_j^d-x_j^s), & \tau \in (0,1)
\end{array}\right.
\eqe
where $\alpha > 0$ is a biasing term such that the projection does not intersect $x_j^s$ or $x_j^d$.  We then have the state transition condition
\eqb{EQ-InFlowCond}
\norm{x_{i\rightarrow \mc{F}_j} -x_i} \leq \epsilon_F
\eqe
which when satisfied gives $\mc{S}_i \gets \textit{SWARMING}$ (line 14, Algorithm \ref{alg:pcp}, and described below).  If condition \eqref{EQ-InFlowCond} is not satisfied, agent $i$ transitions to state $\mc{S}_i \gets \textit{RECONFIGURE}$, applying waypoint controller \eqref{EQ-WayControl} with $x^w \triangleq x_{i\rightarrow \mc{F}_j}$, guaranteeing a reconfiguration, in a shortest path manner, to a point on the line segment defining its assigned flow $\mc{M}_i$.

\begin{figure*}[t]
\centering
\subfloat[]{\label{Fig-SimSnap1} \includegraphics[width=2.35in]{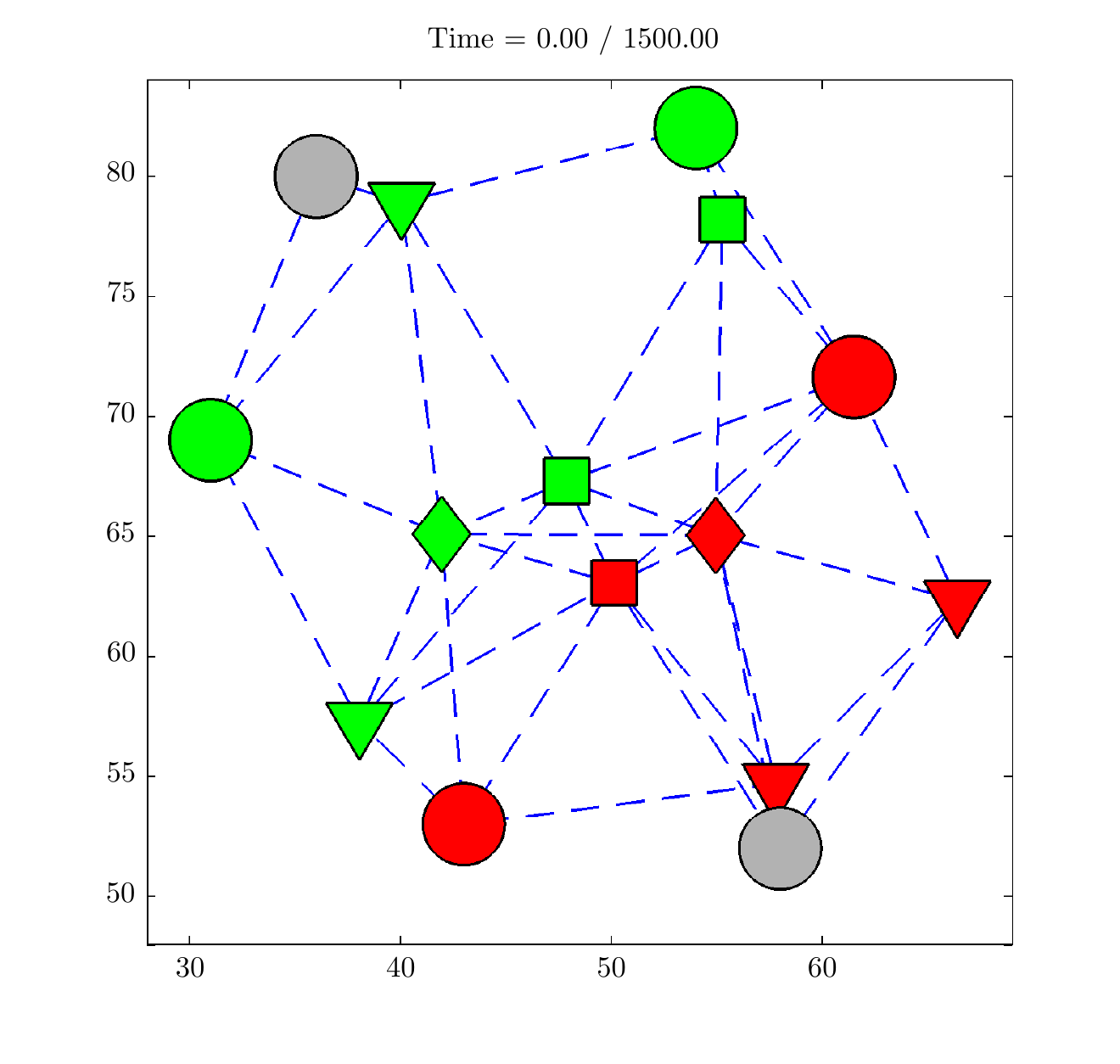}}
\subfloat[]{\label{Fig-SimSnap2} \includegraphics[width=2.35in]{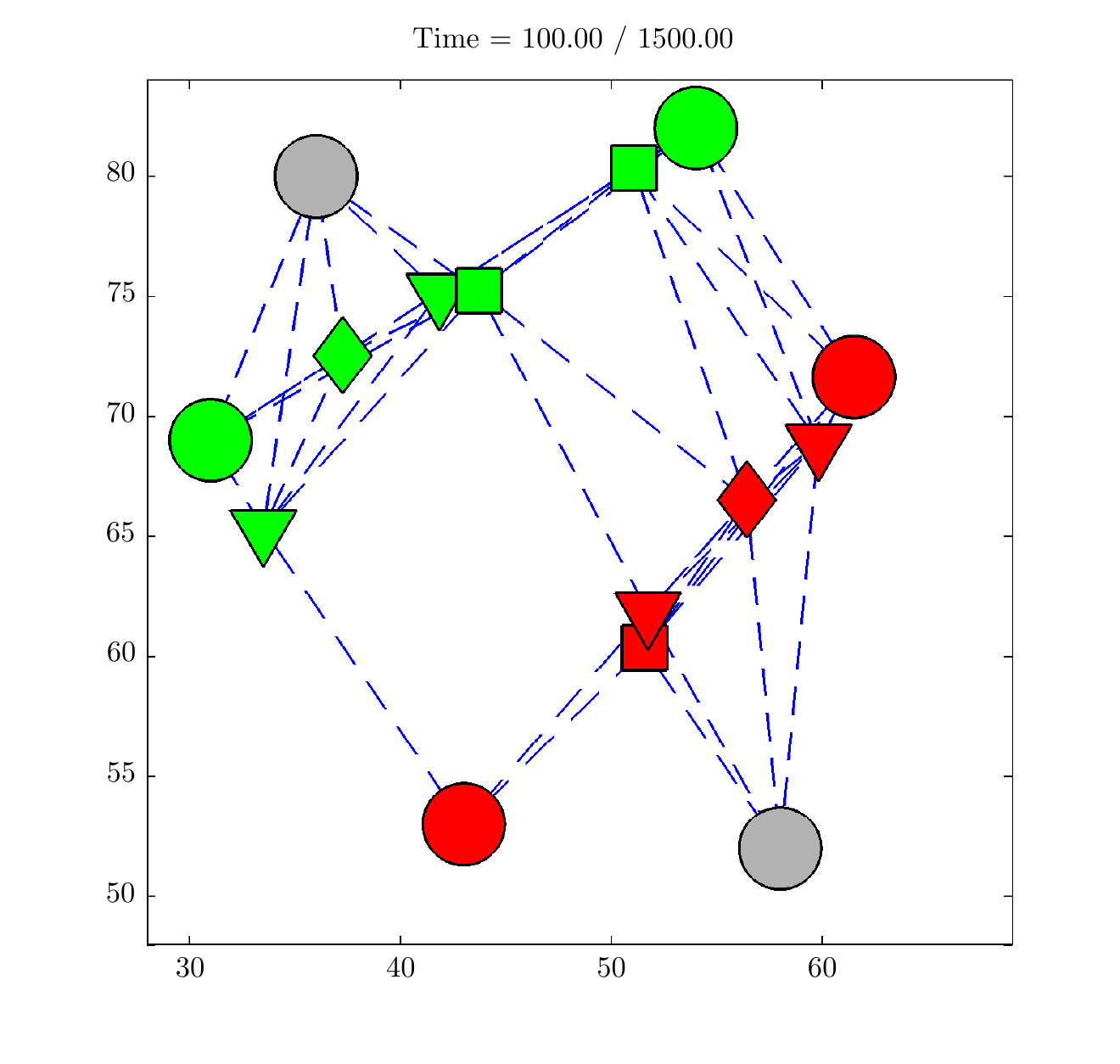}}
\subfloat[]{\label{Fig-SimSnap3} \includegraphics[width=2.35in]{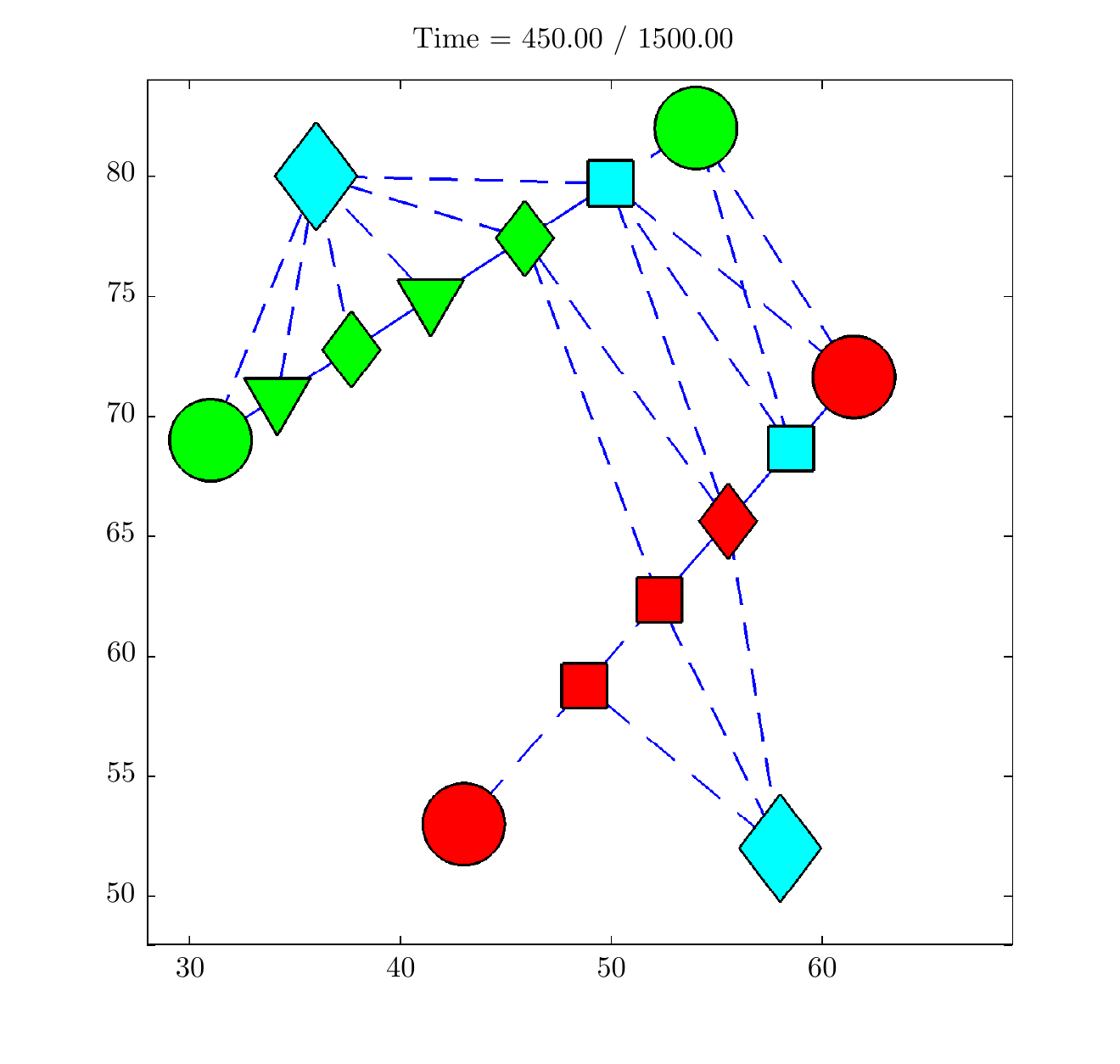}}
\hfill
\subfloat[]{\label{Fig-SimSnap4} \includegraphics[width=2.35in]{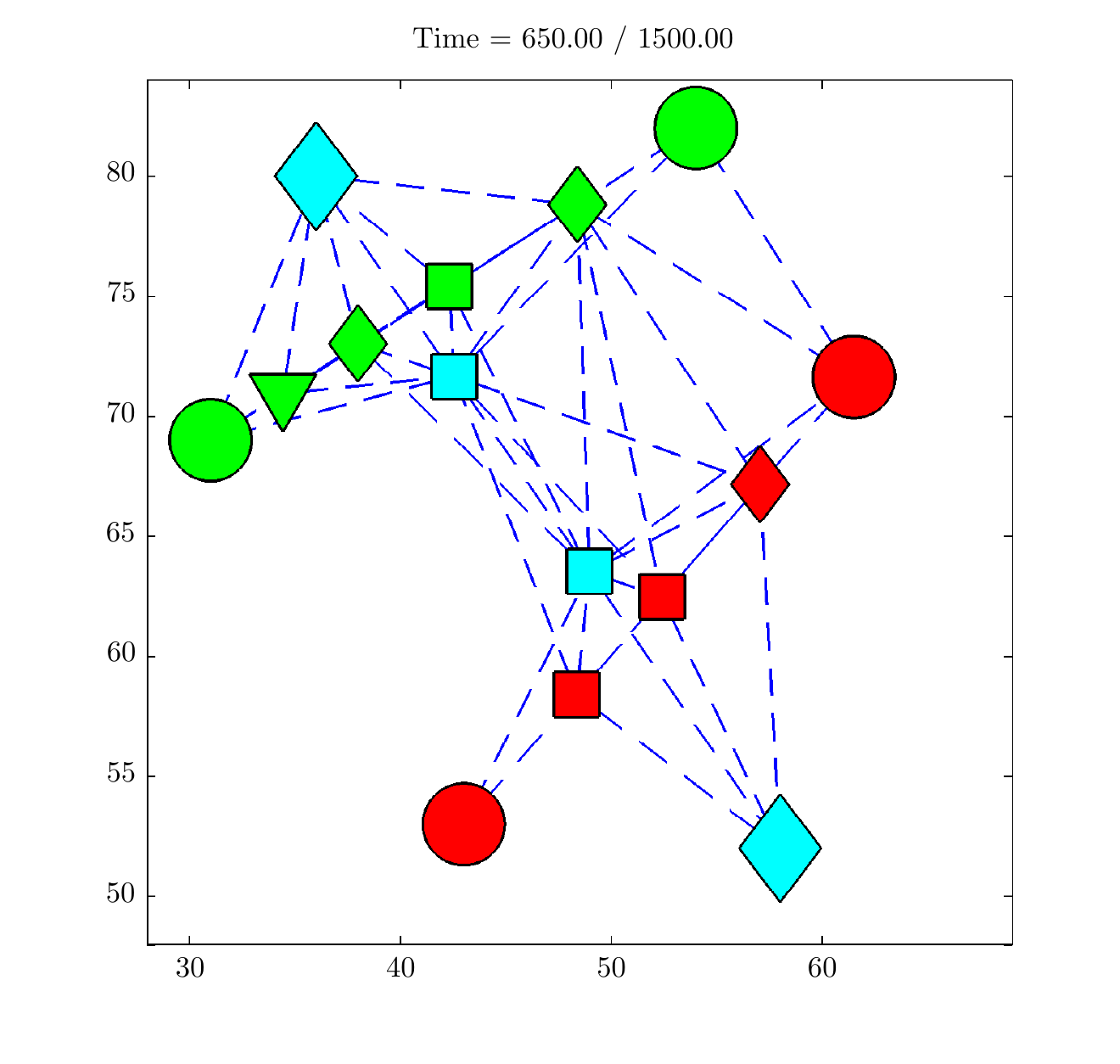}}
\subfloat[]{\label{Fig-SimSnap5} \includegraphics[width=2.35in]{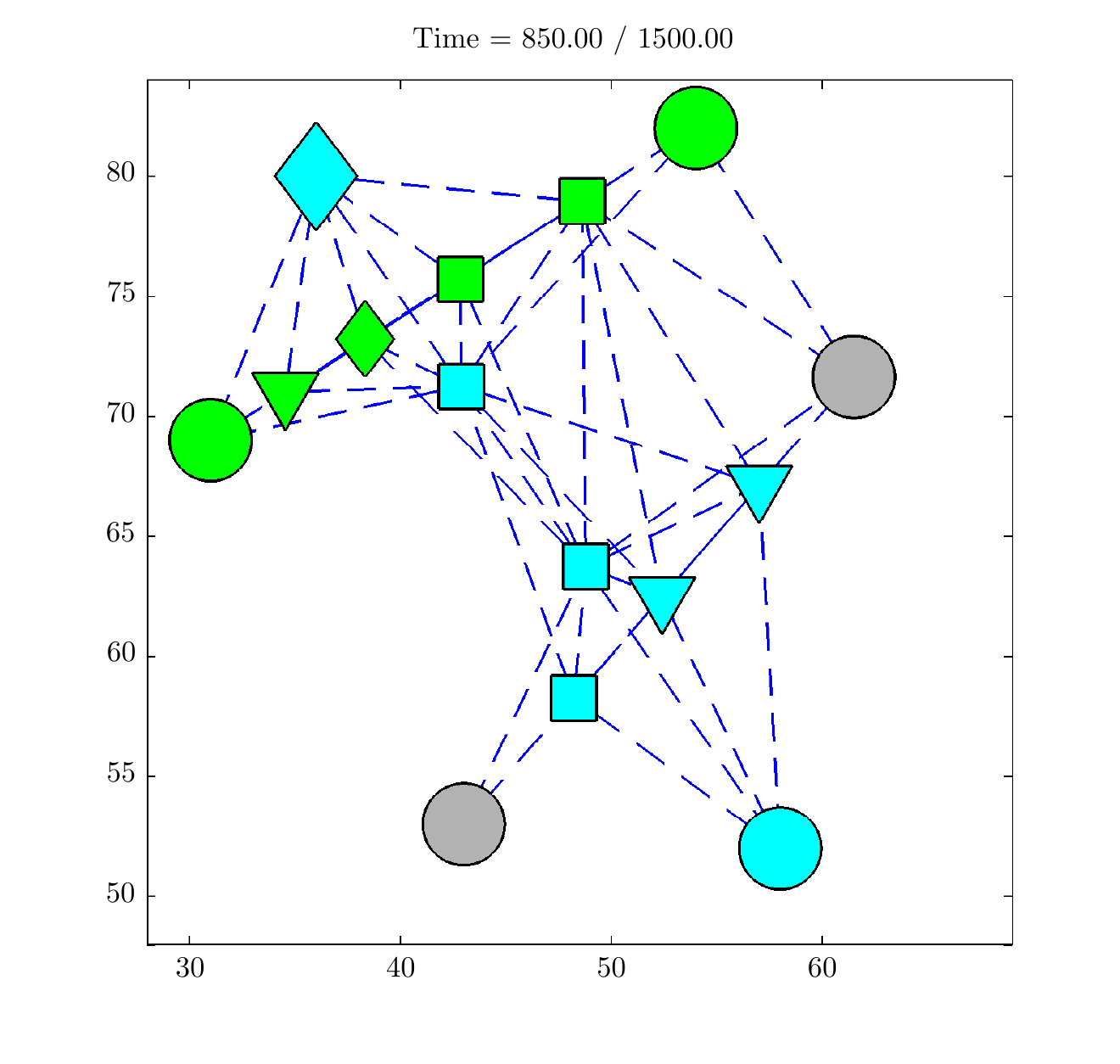}}
\subfloat[]{\label{Fig-SimSnap6} \includegraphics[width=2.35in]{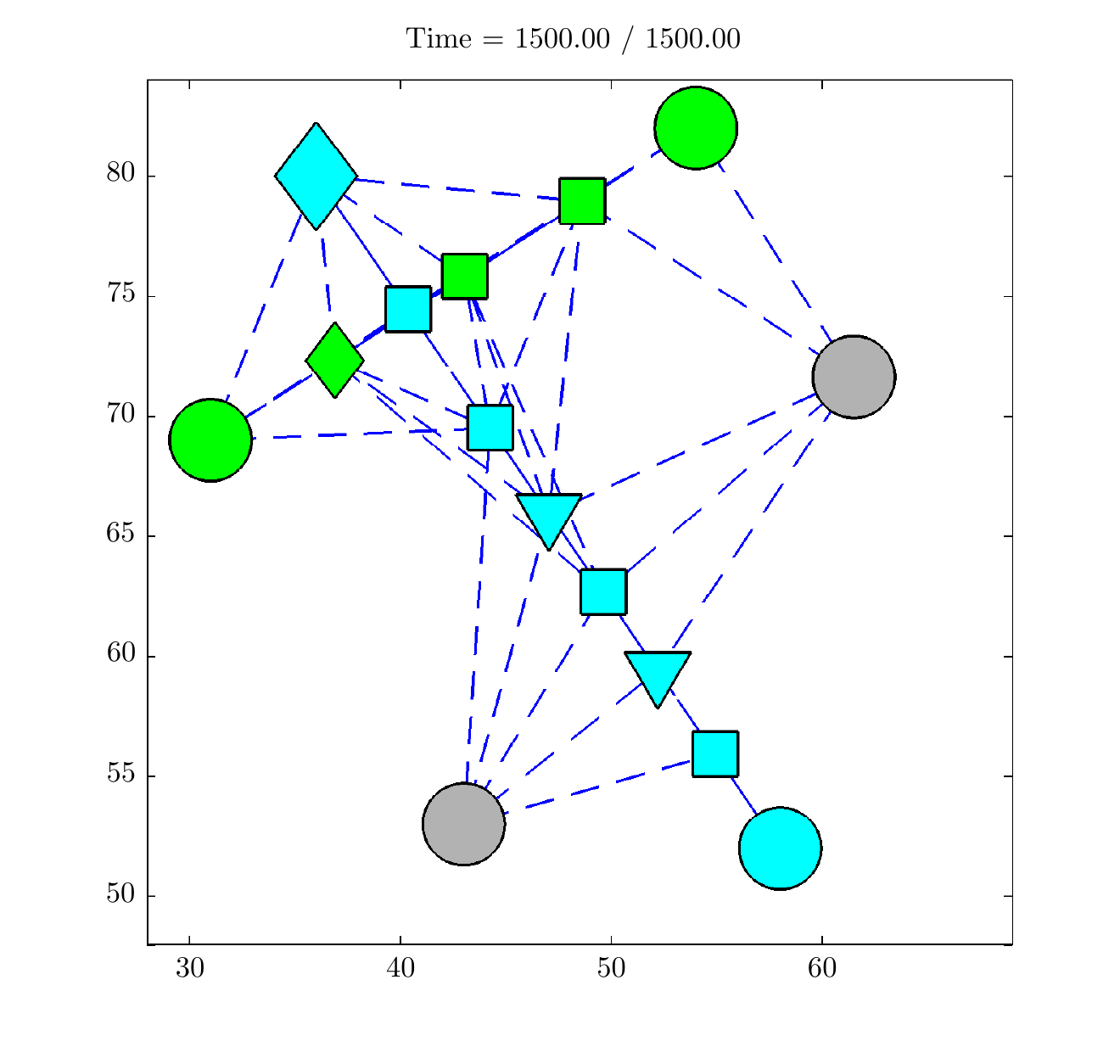}}
\caption{Network progression for the simulated execution described in Section \ref{S-Sim}.  Flow membership is indicated by color, where square nodes are the flow backbone, triangle nodes are redundant with respect to connectivity, and diamond nodes are bridges.  Note that flow $\mc{F}_3$ is initially inactive and becomes active at $t = 450$, while $\mc{F}_2$ is initially active and deactivates at $t = 850$.}
\label{Fig-RouteSwarmSim}
\end{figure*}

Finally, when an agent $i$ is in the swarming state $\mc{S}_i = \textit{SWARMING}$ (lines 18-20, Algorithm \ref{alg:pcp}), after all necessary reconfigurations have been made (either by the ICP via $\mc{C}_i$ or internally by flow alignment), a \emph{dispersive} inter-neighbor controller is applied in order to optimize the assigned flow.  Specifically, each swarming agent $i \in \mc{I}_M$ applies a coordination controller (regardless of bridge status $b_i$):
\eqb{EQ-InFlowDisp}
u_i^o \triangleq -\nabla_{x_i}\sum_{j \in \mc{N}_i^A}\frac{1}{\norm{x_{ij}}^2} - \sum_{j \in \mc{N}_i^S}\nabla_{x_j}\frac{1}{\norm{x_{ji}}^2}
\eqe
where
\eqb{EQ-DispSet1}
\begin{split}
\mc{N}_i^A \triangleq \{&j \in \mc{N}_i \st (\mc{M}_j \cap \mc{M}_i \neq \emptyset) \,\wedge \\
& [(\mc{S}_j = \textit{SWARMING}) \vee (j \in \mc{I}_S)]\}
\end{split}
\eqe
is the set of neighbors that share membership in flow $\mc{M}_i$, and who are either \emph{in flow} and \emph{actively} swarming (i.e.\ by condition \eqref{EQ-InFlowCond}), or are a static source/destination node.  Further, we define
\eqb{EQ-DispSet2}
\mc{N}_i^S \triangleq \{j \in \mc{N}_i^A \st j \in \mc{I}_S\}
\eqe
as the set of static in flow neighbors for which compensation (Remark \ref{Rem-EngComp}, below) must be applied.  Controller \eqref{EQ-InFlowDisp} dictates that mobile flow members disperse equally only with fellow flow members and also with the source/destination nodes of their assigned flow $\mc{M}_i$.

\begin{remark}[Energy compensation]\label{Rem-EngComp}
The inclusion of supplementary control terms for interactions with static neighbors $j \in \mc{N}_i \cap \mc{I}_S$ in \eqref{EQ-InFlowDisp} acts to retain the inter-agent symmetry required for the application of constrained interaction~\cite{Williams:2013bh}, specifically as static agents do not contribute to the system energy.  We refer to this control action as \emph{energy compensation}, an idea that will evolve in future work by Williams and Gasparri to treat systems with asymmetry in sensing, communication, or mobility.
\end{remark}

While dispersive controllers generally yield equilibria in which inter-agent distant is maximized (up to $\rho_2$)~\cite{Dimarogonas:2009}, as each flow is constrained by static source/destination nodes, the dispersion \eqref{EQ-InFlowDisp} generates our desired equidistant intra-flow configuration as formalized below:

\begin{proposition}[Equidistant dispersion]\label{Prop-EquiDisperse}
Consider the application of coordination objective \eqref{EQ-InFlowDisp} to a set of mobile agents $i \in \mc{I}_M$ within the context of interaction controller \eqref{EQ-ConIntControls}, each sharing membership to a flow $k \in \mc{I}_F$, i.e.\ $\mc{M}_i = k, \fa i$.  It follows that at equilibrium the agents are configured such that the \emph{equidistant spacing} condition
\eqb{EQ-EquiCond}
\norm{x_{ij}} \rightarrow \frac{\norm{x_k^d-x_k^s}}{\abs{\mc{V}_F^k}-2}, \quad \fa i \st j \in \mc{N}_i^F
\eqe
holds asymptotically over flow $k$.
\end{proposition}

A formal proof is beyond the scope of this work\footnote{Informally, an energy balancing argument establishes the result.}, however note that our controllers operate using \emph{only} inter-agent distance, an advancement beyond related works such as \cite{Goldenberg04}.

\section{Simulation Results} \label{S-Sim}

\begin{figure}[t]
\centering
\includegraphics[width=3.0in]{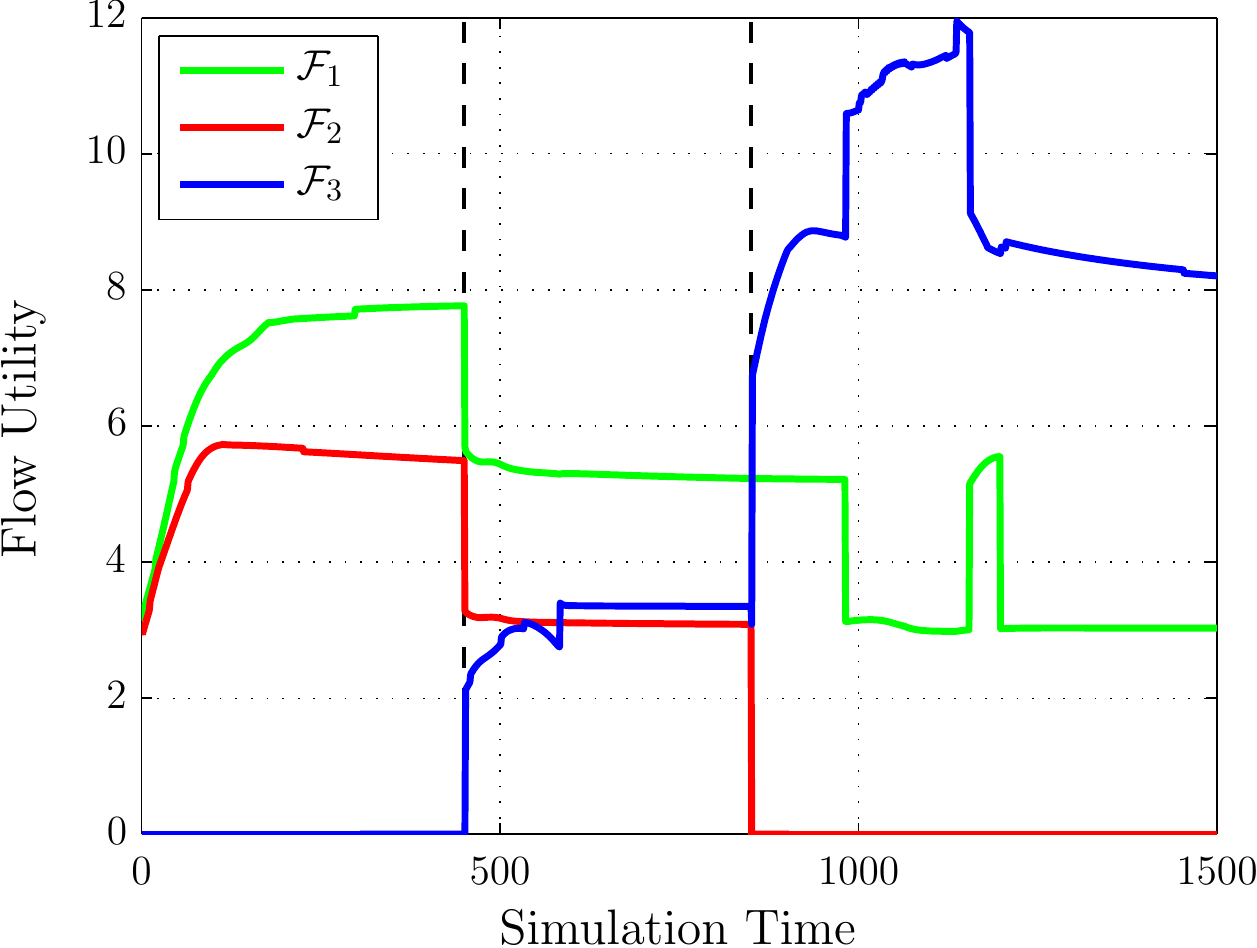}
\caption{Flow utilities for the simulated execution described in Section \ref{S-Sim}.}
\label{Fig-SimFlowUtility}
\end{figure}

In this section, we present a simulated execution of our described INSPIRE proof-of-concept, Route Swarm.  Consider a system operating over a workspace in $\mbb{R}^2$, having $n = 15$ total agents, $m = 9$ of which are mobile and $s = 6$ of which are static information source/destinations.  Assume we have $f=3$ flows (green, red, and blue indicate flow membership), with the initial system configuration depicted as in \figurename \ref{Fig-SimSnap1} (notice that $\mbb{G}$ is initially connected), with the system dynamics shown in \figurename \ref{Fig-SimSnap2} through \ref{Fig-SimSnap6}.  We simulate a scenario in which $\mc{F}_3$ is initially \emph{inactive} (gray), allowing the ICP to optimize agent allocation over only $f=2$ flows, as in \figurename \ref{Fig-SimSnap2} to \ref{Fig-SimSnap3}.  By \figurename \ref{Fig-SimSnap3}, flows $\mc{F}_1$ and $\mc{F}_2$ have been assigned an evenly distributed allocation of mobile agents, where the PCP has provided equidistant agent spacing for each flow.  At this same time (650 time steps), the flow $\mc{F}_3$ activates, initiating a reconfiguration by the ICP to optimize the newly added flow, as in \figurename \ref{Fig-SimSnap4}, noting that initially in \figurename \ref{Fig-SimSnap3}, $\mc{F}_3$ is poorly served by the network configuration.  Finally, in \figurename \ref{Fig-SimSnap5}, flow $\mc{F}_2$ is deactivated, forcing another reconfiguration yielding the equilibrium shown in \figurename \ref{Fig-SimSnap6}.   The per-flow utility over the simulation, given for a flow $i \in \mc{I}_F$ by $\sum_{(j,k) \in \mc{E} \st i \in \mc{M}_j \cap \mc{M}_k}w_{ij}$ (the sum of the link utilities associated with each flow), is depicted in \figurename \ref{Fig-SimFlowUtility}.  Finally, to better illustrate the dynamics of our proposed algorithms, we direct the reader to \mbox{\url{http://anrg.usc.edu/www/Downloads}} for the associated simulation video.

\begin{remark}[Dynamic vs.\ static]
The optimizations proposed in this work are advantageous in terms of dynamic information flow needs and changing system objectives, when compared to static solutions.  On flow switches, static placements fail to fulfill the information flow needs of the altered system configuration.  Additionally, our methods allow for dynamics in $\mc{I}_M$ itself, as the ICP can adaptively reconfigure the system to utilize the available agents across the network flows.
\end{remark}

\section{Conclusion} \label{sec:concl}
In this paper, we illustrated a novel hybrid architecture for command, control, and coordination of networked robots for sensing and information routing applications,  called INSPIRE (for INformation and Sensing driven PhysIcally REconfigurable robotic network). INSPIRE provides of two control levels, namely  Information Control Plane and Physical Control Plane, so that a feedback between information and sensing needs and robotic configuration is established. An instantiation was provided as a proof of concept where a mobile robotic network is dynamically reconfigured to ensure high quality routes between static wireless nodes, which act as source/destination pairs for information flow. Future work will be focused on the validation of the proposed architecture in a real-world scenario having mobile robotic interaction with a sensor network testbed.

\bibliographystyle{IEEEtran}
\bibliography{biblio}

\end{document}